# On electron imaging of small molecules, Part I: relative roles of multiple scattering and Fresnel diffraction

T.E. Gureyev[*,1,2,3,4], H.M. Quiney[1], A. Kozlov[1] and L.J. Allen[1]

[1)] *ARC Centre in Advanced Molecular Imaging, School of Physics, The University of Melbourne, Parkville 3010, Australia*

[2)] *School of Physics and Astronomy, Monash University, Clayton 3800, Australia*

[3)] *School of Science and Technology, University of New England, Armidale 2351, Australia*

[4)] *Faculty of Health Science, The University of Sydney, Sydney 2006, Australia*

**Abstract**

The relative roles of multiple electron scattering and in-molecule free-space propagation in transmission electron microscopy of small molecules are discussed. It is argued that while multiple scattering tends to have only a moderate effect in this case, the in-molecule Fresnel diffraction is likely to be significant due to the shallow depth of focus under the relevant experimental conditions. As a consequence, diffraction tomography based on the first Born or first Rytov approximation represents a more suitable method for the reconstruction of three-dimensional distribution of the electrostatic potential in this context, compared to conventional computed tomography which is intrinsically based on the projection approximation. A simplified method for diffraction tomography is proposed and tested on numerically simulated examples.

**1. Introduction**

It is well known that in transmission electron microscopy (TEM) multiple scattering effects are significant for all but the thinnest of specimens and cannot be ignored in the process of atomic structure determination, especially from crystalline specimens [1]. However, in areas like cryogenic electron microscopy (cryo-EM) multiple scattering is typically ignored, on the basis that for most biological molecules of interest the samples under investigation are non-crystalline and sufficiently thin [2-4]. Computed tomography (CT) [5] is often used for the reconstruction of a three-dimensional (3D) structure from projection images collected at different orientations (angular positions) of the sample, or from images of multiple copies of identical samples in random orientations. Conventional CT algorithms assume not only the absence of multiple scattering in the data acquisition process, but also the applicability of the projection approximation [6] on which the CT is fundamentally based [5]. The latter assumption means that the free-space propagation effects, i.e. the Fresnel diffraction inside the sample, must also be negligible. The projection approximation is associated with the flatness of the corresponding Ewald sphere and the sufficient depth of focus of the imaging setup, which is assumed to exceed the thickness of the sample. However, in high-resolution TEM the latter assumption usually does not hold, since the corresponding depth of focus is quite shallow (being of the order of the thickness of a single atomic layer) and the curvature of the Ewald sphere cannot be neglected. One prominent practical consequence of this fact can usually be easily verified: if the sample is illuminated by a plane wave, then the TEM projection image changes in a non-trivial manner when the sample is rotated by 180 degrees around an axis perpendicular to the direction of the incident wave (see an example in Fig. 1). Such an effect directly contradicts the validity of the projection approximation, because, in that approximation, the image is formed by integrating the electrostatic potential along a straight line, and the result cannot depend on the direction of the integration.

---

[*] Corresponding author. *E-mail address*: timur.gureyev@unimelb.edu.au (T.E. Gureyev).

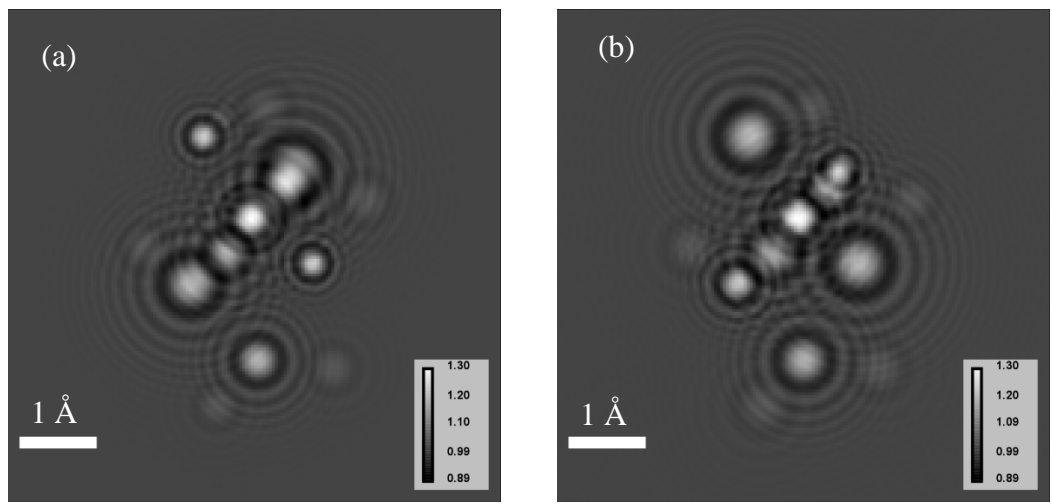

Fig. 1. Defocused images of the same molecule of aspartate, $C_4H_7NO_4$, simulated with monochromatic, $E = 200$ keV, plane electron wave illumination, at two different orientations of the molecule differing by 180 degrees rotation around the vertical ($y$) axis. Image (b) is also mirror-reflected with respect to the central vertical axis for display purposes. Differing image contrast produced by the same atoms in the two images is attributable to different distances between these atoms and the image plane in the two orientations.

One might hope that this problem can be removed or substantially alleviated if one applies a so-called contrast transfer function (CTF) correction procedure to the TEM images prior to CT reconstruction [8]. However, at least in its conventional forms, the CTF also assumes the projection approximation for the propagation of the wave through an "infinitely thin" sample with *a priori* unknown structure. The latter approximation again means that, after the CTF correction, the images of the sample in two different orientations differing by 180 degree rotation are going to be identical, except for the trivial mirror reflection with respect to the rotation axis. In order to carry out the "correct" CTF compensation, each individual atom in the molecule would need to have a different degree of CTF compensation (refocusing) applied, in accordance with the distance between the atom and the image plane. Naturally, such a procedure generally requires *a priori* knowledge of each atom position along the optical axis, which is usually not available. However, it is also known that in optical setups with shallow depth of focus and weakly scattering samples, a different method termed "diffraction tomography" (DT) [9-11] can be used to properly account for the Fresnel diffraction in the sample and correctly reconstruct the 3D distribution of the complex refractive index, or, equivalently, the spatial distribution of the electrostatic potential in electron imaging. The DT approach can be based on the first Rytov or first Born approximation, instead of the projection approximation utilized in the conventional CT. As a consequence, the DT projections of the sample in opposite orientations can be different, because the method takes into account the different distances between atomic planes and the image plane in each orientation. Note, however, that while DT takes into account the free-space propagation, it does not fully account for multiple scattering.

The well-known multislice approach [12] correctly takes into account both of the effects discussed above, i.e. the Fresnel diffraction and multiple scattering inside a sample. This method is commonly used for "forward" TEM simulations [13]. It is, however, quite challenging to utilize this technique for the solution of the corresponding "inverse" problem, i.e. for the reconstruction of the 3D structure of the sample from TEM projections, with the first such results reported only very recently [14]. In the present work, we have performed some simulations of TEM images of single biological molecules, namely aspartate, lysozyme, lasso peptide and others (examples are presented below) using a well-known, freely available software package TEMSIM [15] developed by E.J. Kirkland on the basis of the multislice method [13]. Our simulations presented below consistently indicated that for TEM images of single biological molecules, obtained with a plane monochromatic electron waves with energies of $E \sim 200 - 300$ keV, the multiple scattering effects (both within an atom and between different atoms) were relatively weak, producing an effect of no more than a few percent. On the other hand, the in-molecule free-space propagation effects were significant, in the sense that neglecting them resulted in errors up to tens of percent.

**2. Multiple scattering and Fresnel diffraction in projection images and CT reconstructions of biological molecules**

As a first example, we compared the result of a full multislice-based projection image through a lysozyme molecule (Fig. 2(a)) with a similar one obtained on the basis of the projection approximation (which ignores both the multiple scattering between different atoms and the in-molecule Fresnel diffraction, but not the multiple scattering within an atom), the average error in the image intensities was 3.4% and the maximum error was 58% (Fig. 2(b)). The error was not distributed evenly across the whole image, but was concentrated predominantly around atoms located further from the image plane. On the other hand, when we simulated a similar image of the same molecule using the multislice approximation for each single atom of the molecule separately and then incoherently adding together all the images of the individual atoms, thus ignoring multiple scattering but not the Fresnel diffraction, the average error in the image intensities was 0.6% and the maximum error was 10% (Fig. 2(c)).

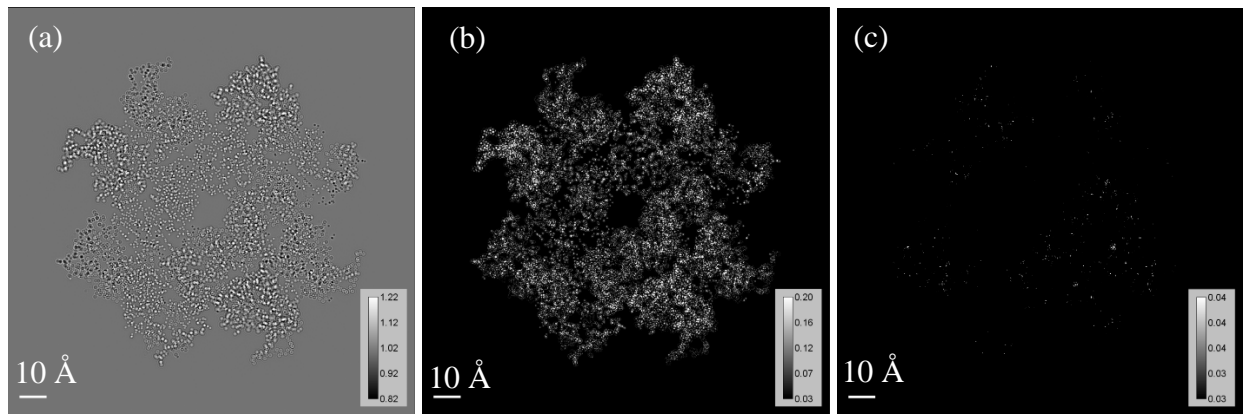

Fig. 2. Simulated images of lysozyme molecule corresponding to a monochromatic plane electron wave with energy $E = 200$ keV. Image size is $150 \times 150$ Å$^2$. (a) Full multislice projection refocused back to the central transverse, $(x, y)$, plane; (b) error map of the projection approximation (free-space propagation and multiple scattering both ignored); (c) error map of the composite image obtained by multislice imaging of each single atom separately (only multiple scattering ignored). Maps (b) and (c) have been thresholded from below at 3% error for display purpose.

We have also verified that the large error associated with the use of the projection approximation in TEM indirectly leads to significant artefacts in the 3D map of the electrostatic potential reconstructed using conventional CT algorithms. In one such test, the simulated input data for the CT reconstruction consisted of multislice projections of lysozyme molecule, CTF-corrected by backpropagating the exit-plane wave to the central plane (containing the axis of rotation), and calculated for 1800 rotational positions of the molecule with respect to the $y$ axis over the span of 360 degrees with a step of 0.2 degrees. One CT-reconstructed axial, i.e. $(x, z)$, slice through the reconstructed 3D distribution of the electrostatic potential is shown in Fig. 3(a). The reconstruction contains strong artefacts: for example, it is easy to see that some reconstructed "atoms" at the central region of the slice display dark (negative) contrast, while others display light (positive) contrast. In this case, the artefacts in the CT-reconstructed electrostatic potential were not caused by errors in the simulated multislice projections, but rather by the fact that the conventional CT reconstruction itself is intrinsically based on the projection approximation, i.e. on the inherent notion that the input data was formed by integrating the electrostatic potential along straight lines. In order to verify that this was indeed the main cause for the reconstruction errors, we have checked that, under the same simulation conditions, when the forward projections were calculated according to the projection approximation (instead of the multislice method), the reconstructed distribution did not have any significant artefacts (Fig. 3(b)).

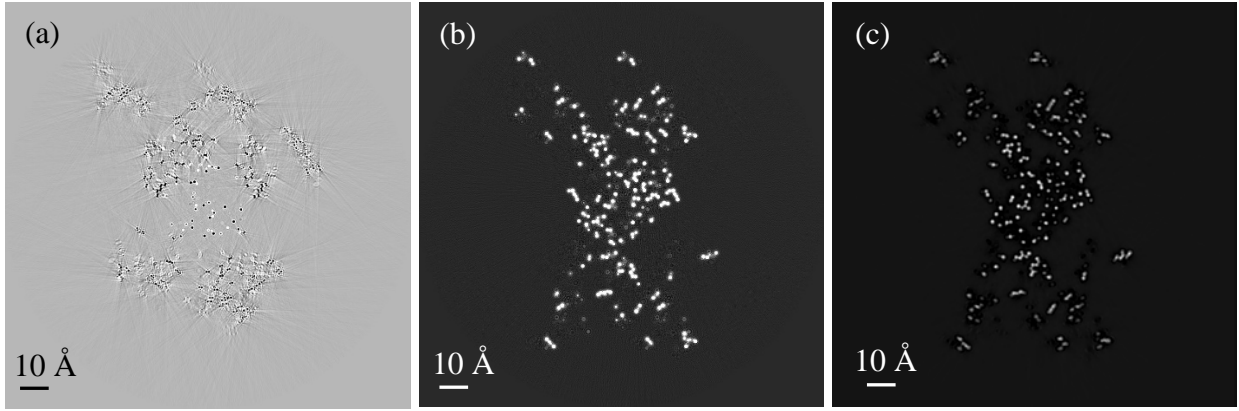

Fig. 3. Reconstructed axial CT slice through a Lysozyme molecule obtained from monochromatic plane electron wave projections with energy $E = 200$ keV. (a) CT slice obtained from 1800 CTF-corrected full multislice projections; (b) same CT slice as in (a), but reconstructed from straight-ray projections of the electrostatic potential; (c) same slice obtained from 1800 full multislice projections using a simplified DT approach (see details in Section 3).

The above considerations suggest that, for the TEM imaging of single molecules, a DT-type reconstruction that ignores multiple scattering, but takes in-molecule free-space propagation into account, is likely to perform well [7]. One such approach is described in the next section. While it may be possible to find much of the material presented in that section in previous publications (see e.g. [7-11]), we have not been able to find a convenient source where similar results would have been presented in a self-contained form relevant for TEM. Here, using the conventional CTF theory [12] as a starting point, we first derive an expression for TEM image intensity as an incoherent sum of the first Born approximations for scattered intensities by atoms located in each plane orthogonal to the incident beam. We then show how this expression can be used in a DT-type approach [11] for 3D reconstruction of the sample from TEM images collected at multiple angular orientations of the sample. Figure 3(c) shows an example of a DT-type reconstruction, using a simplified Transport of Intensity equation based approach, as presented at the end of Section 3. This result clearly displays much better quality of the reconstruction of the molecule, with much fewer artefacts, compared to the conventional CT reconstruction obtained from the same multislice projections (Fig. 3(a)).

## 3. Diffraction tomography based on first Born approximation

Consider an imaging setup with a monochromatic plane wave $I_{in}^{1/2} \exp(i2\pi k z)$ illuminating a weakly scattering object, where $k = 1/\lambda$ is the wave number, $I_{in} = const$ is the intensity of the wave and $\mathbf{r} \equiv (x, y, z)$ is the Cartesian coordinate system in 3D space. The complex amplitude $U(\mathbf{r})$ of the wave inside the object satisfies the stationary wave equation $\nabla^2 U(\mathbf{r}) + 4\pi^2 n^2(\mathbf{r}) k^2 U(\mathbf{r}) = 0$, where $n(\mathbf{r})$ is the refractive index. In the case of electron microscopy, one has $n(\mathbf{r}) \cong 1 + V(\mathbf{r})/(2E)$, where $V(\mathbf{r}) \geq 0$ is the electrostatic potential, $E$ is the accelerating voltage and $V(\mathbf{r})/(2E)$ is typically much less than 1 (see e.g. [16]). We consider the problem of reconstruction of the 3D distribution of the electrostatic potential from the intensity of transmitted waves measured at some distance(s) from the object, for a set of different rotational positions of the object.

Recall that, when the object is thin compared to the depth of focus and is weakly scattering, the first Born approximation to the intensity of the projection image collected at a position $z$ downstream from the object along the optic axis can be expressed as [12]:

$$(\mathbf{F}_2 I_{thin})(\mathbf{q}_\perp, z) / I_{in} = \delta(\mathbf{q}_\perp) + 2\sin[\pi\lambda z q_\perp^2](\mathbf{F}_2 \varphi)(\mathbf{q}_\perp), \qquad (1)$$

where $\varphi(\mathbf{r}_\perp) = [\pi/(\lambda E)]\int_{-\delta z}^{0} V(\mathbf{r}_\perp, z')dz'$, $\mathbf{r}_\perp \equiv (x,y)$, the object is contained within a thin slab $[-\delta z, 0]$, $(\mathbf{F}_2 f)(\mathbf{q}_\perp) \equiv \iint \exp[-i2\pi\mathbf{q}_\perp \mathbf{r}_\perp] f(\mathbf{r}_\perp)d\mathbf{r}_\perp$ is the 2D Fourier transform, $\mathbf{q}_\perp \equiv (q_x, q_y)$ and $q_\perp \equiv |\mathbf{q}_\perp|$. A "thick" object located in a slab $[z_0, 0]$, $z_0 < 0$, can be split into a sufficiently large number, $M$, of thin slices, with each slice contained in a thin slab $[z_m, z_{m+1}]$, $z_{m+1} = z_m + \delta z$, $m=0,1,\ldots,M$-1, $z_M = 0$ and $\delta z = |z_0|/M$. Assuming that each slice diffracts only weakly and incoherently with respect to all other slices, and that the incident plane wave arrives unperturbed at each transverse slice (which can be referred to as a "kinematical approximation" [12]), we can represent the total diffracted intensity as an incoherent sum of intensities diffracted from individual thin slices:

$$(\mathbf{F}_2 I)(\mathbf{q}_\perp, z) / I_{in} = \delta(\mathbf{q}_\perp) + 2\sum_{m=0}^{M-1} \sin[\pi\lambda(z-z_m)q_\perp^2](\mathbf{F}_2 \varphi_m)(\mathbf{q}_\perp), \qquad (2)$$

where $\varphi_m(\mathbf{q}_\perp) \equiv [\pi/(\lambda E)]\int_{z_m}^{z_{m+1}} V(\mathbf{q}_\perp, z')dz'$. Note that the incoherent approximation is natural in this context, because the coherent interference terms are of the second order with respect to the implicit small parameter in the Born series, and hence can be neglected. When the thickness of each slice is very small, we can approximate $\varphi_m(\mathbf{q}_\perp) \cong [\pi/(\lambda E)] V(\mathbf{q}_\perp, z_m)\delta z$. Substituting this into Eq. (2), and letting the slice thickness go to zero, $\delta z \to 0$, while simultaneously letting $M \to \infty$, so that the total thickness of the slab $|z_0| = M\delta z$ remains constant, we obtain:

$$(\mathbf{F}_2 I)(\mathbf{q}_\perp, z) / I_{in} = \delta(\mathbf{q}_\perp) + [2\pi/(\lambda E)] \int \sin[\pi\lambda(z-z')q_\perp^2](\mathbf{F}_2 V)(\mathbf{q}_\perp, z') dz', \tag{3}$$

where we have also extended the limits of the integral over $z'$ to $\pm\infty$ by formally assuming that $V(\mathbf{r}) = 0$ outside the slab $[z_0, 0]$. Expressing the sine function under the integral sign in Eq. (3) via a difference of two complex exponents and introducing the contrast function, $K(\mathbf{r}_\perp, z) \equiv 1 - I(\mathbf{r}_\perp, z)/I_{in}$,, we arrive at

$$\begin{aligned}&(\mathbf{F}_2 K)(\mathbf{q}_\perp, z) \\ &= [i\pi/(\lambda E)][\exp(i\pi\lambda z q_\perp^2)(\mathbf{F}_3 V)(\mathbf{q}_\perp, (\lambda/2)q_\perp^2) - \exp(-i\pi\lambda z q_\perp^2)(\mathbf{F}_3 V)(\mathbf{q}_\perp, -(\lambda/2)q_\perp^2)],\end{aligned} \tag{4}$$

where $(\mathbf{F}_3 V)(\mathbf{q}_\perp, q_z) \equiv \iint \exp[-i2\pi(\mathbf{r}_\perp \mathbf{q}_\perp + z q_z)] V(\mathbf{r}_\perp, z) d\mathbf{r}_\perp dz$ is the 3D Fourier transform of the electrostatic potential.

This equation is reminiscent of the Fourier slice theorem which is used as a basis for object reconstruction in CT [5,9]. In order to transform Eq. (4) into the conventional Fourier slice theorem (for the case of a weak phase-contrast object) one would need to replace the terms $\pm(\lambda/2)q_\perp^2$ by zero, which corresponds to the projection approximation [10]. Note that, when $k_\perp^2 \ll k^2$ according to the paraxial conditions, we have $k_\perp^2/(2k) \cong k - [k^2 - k_\perp^2]^{1/2} = k - k_z$, where $k_z^2 = k^2 - k_\perp^2$. Therefore, the vector $(\mathbf{k}_\perp, -(\lambda/2)\mathbf{k}_\perp^2) \cong (\mathbf{k}_\perp, k_z - k)$ describes a (small) difference between the scattered wave vector $\mathbf{k}_1 \equiv (\mathbf{k}_\perp, k_z)$ lying on the Ewald sphere and the incident wave vector $\mathbf{k}_0 = (0, 0, k)$: $\mathbf{k}_1 - \mathbf{k}_0 = (\mathbf{k}_\perp, k_z - k)$. Note that, unlike X-ray CT, in high-resolution electron microscopy the curvature of the Ewald sphere generally cannot be neglected even in the paraxial regime, i.e. it is incorrect to replace $\pm(\lambda/2)q_\perp^2$ with zero in Eq. (4). Indeed, for such an approximation to be accurate, it would be required that $(\lambda/2)|z_0|(q_{x,\max}^2 + q_{y,\max}^2) \ll 1$ or $N_{\min}^F \equiv a^2/(\lambda|z_0|) \gg 1$, where $a \equiv q_{x,\max}^{-1} = q_{y,\max}^{-1}$ is the size of the smallest resolvable feature in the object and $|z_0|$ is the extent of the object in the $z$ direction. If the desired spatial resolution is close to 1 Å, the size of the object (e.g. a biological molecule) is around 100 Å and the wavelength is 0.025 Å (corresponding to 200 keV electrons), then $N_{\min}^F = 0.4 < 1$, and hence the validity condition for the projection approximation is not satisfied and the curvature of the Ewald sphere needs to be taken into account.

We would like to solve Eq. (4) with respect to the potential *V*. Generally it is impossible to retrieve the electrostatic potential from Eq. (4) alone, as it contains two unknown values, $\exp(i\pi\lambda z q_\perp^2)(\mathbf{F}_3 V)(\mathbf{q}_\perp,(\lambda/2)q_\perp^2)$ and $\exp(-i\pi\lambda z q_\perp^2)(\mathbf{F}_3 V)(\mathbf{q}_\perp,-(\lambda/2)q_\perp^2)$. Note however that if the contrast function for the object rotated by 180 degrees around the *y* axis, $K_\pi(\mathbf{q}_\perp,z)$, is also available, for such a projection we have from Eq. (3):

$$\begin{aligned}(\mathbf{F}_2 K_\pi)(\mathbf{q}_\perp^-,z) &= [2\pi/(\lambda E)]\int \sin[\pi\lambda(z'-z)q_\perp^2](\mathbf{F}_2 V)(\mathbf{q}_\perp,-z')dz' \\ &= [2\pi/(\lambda E)]\int \sin[\pi\lambda(-z'-z)q_\perp^2](\mathbf{F}_2 V)(\mathbf{q}_\perp,z')dz' \\ &= [i\pi/(\lambda E)][\exp(i\pi\lambda z q_\perp^2)(\mathbf{F}_3 V)(\mathbf{q}_\perp,-(\lambda/2)q_\perp^2) - \exp(-i\pi\lambda z q_\perp^2)(\mathbf{F}_3 V)(\mathbf{q}_\perp,(\lambda/2)q_\perp^2)],\end{aligned} \quad (5)$$

where $\mathbf{q}_\perp^- \equiv (-q_x,q_y)$ is the mirror-reflection of the vector $\mathbf{q}_\perp = (q_x,q_y)$ with respect to the rotation axis. The system of two linear equations (4)-(5) can be easily solved with respect to $(\mathbf{F}_3 V)(\mathbf{q}_\perp,-(\lambda/2)q_\perp^2)$:

$$(\mathbf{F}_3 V)(\mathbf{q}_\perp,-(\lambda/2)q_\perp^2) = \frac{-\lambda E}{2\pi}\left[\frac{\exp(-i\pi\lambda z q_\perp^2)(\mathbf{F}_2 K)(\mathbf{q}_\perp,z)+\exp(i\pi\lambda z q_\perp^2)(\mathbf{F}_2 K_\pi)(\mathbf{q}_\perp^-,z)}{\sin(2\pi\lambda z q_\perp^2)}\right]. \quad (6)$$

This formula represents a variant of the diffraction Fourier slice theorem [10,11] in the first Born approximation. In a general situation, as the denominator of the right-hand side of Eq. (6) may be equal to zero at some points, it should be regularized, e.g. using the conventional Tikhonov regularization [17]. Also, if projection data is available for more than one defocus distance at each rotational position, this too can be effectively used for regularization of Eq. (6) [7]. Finally, when projections for a set of rotational positions of the sample are made available from the experimental measurements (together with the corresponding projections differing by 180 degree rotation around the *y* axis), such that the paraboloids $(\mathbf{q}_\perp,-(\lambda/2)q_\perp^2)$ fill in the whole reciprocal 3D sphere, the spatial distribution of the electrostatic potential, $V(\mathbf{r})$, in the sample can be obtained by 3D Fourier inversion of $(\mathbf{F}_3 V)(\mathbf{q})$. In practice, of course, the discrete set of measured angular positions and the detector pixel density in each such position should satisfy the usual Nyquist sampling conditions [5] in order for this inversion to be accurate. Note that if the projection approximation is also valid, then $K_\pi(\mathbf{q}_\perp^-,z) = K(\mathbf{q}_\perp,z)$, as discussed in the Introduction, and also $\lambda z q_\perp^2 \ll 1$, $\exp(\pm i\pi\lambda z q_\perp^2) \cong 1 \pm i\pi\lambda z q_\perp^2$ and $\sin(2\pi\lambda z q_\perp^2) \cong 2\pi\lambda z q_\perp^2$. It is easy to verify that under these approximations, Eq. (6) takes the form of the 2D Fourier transform of the Transport of Intensity equation (TIE) in the case of a phase object, $I(\mathbf{r}_\perp, z) = I_{in}\{1 - [\lambda z/(2\pi)]\nabla_\perp^2 \varphi(\mathbf{r}_\perp)\}$ [6], where $\varphi(\mathbf{r}_\perp) = [\pi/(\lambda E)]\int V(\mathbf{r}_\perp, z)dz$.

If the molecule is small and the defocus distance is sufficiently short too, it may be possible to approximate $\sin[\pi\lambda(z-z')q_\perp^2] \cong \pi\lambda(z-z')q_\perp^2$ in Eq. (3). In this case, after making a similar approximation in Eq. (5), adding the results together and taking the 2D inverse Fourier transform, we obtain

$$K(\mathbf{r}_\perp, z) + K_\pi(\mathbf{r}_\perp^-, z) = (z/E)\nabla_\perp^2 \int V(\mathbf{r}_\perp, z)dz. \tag{7}$$

Equation (7) also represents a form of the TIE, since it can be re-written as $[I(\mathbf{r}_\perp, z) + I_\pi(\mathbf{r}_\perp^-, z)]/2 = I_{in}\{1 - [\lambda z/(2\pi)]\nabla_\perp^2 \varphi(\mathbf{r}_\perp)\}$. It allows one to perform a simple DT-type reconstruction from defocused projections acquired over 360 degree rotation of the sample around the $y$ axis. On the first step ("phase retrieval"), the symmetrized contrast function $\tilde{K}_\theta(\mathbf{r}_\perp, z) = [K_\theta(\mathbf{r}_\perp, z) + K_{\theta+\pi}(\mathbf{r}_\perp^-, z)]/2$, where $\theta$ is the sample rotation angle, can be used as input for the TIE inversion formula, $\varphi_\theta(\mathbf{r}_\perp) = [2\pi/(\lambda z)]\nabla_\perp^{-2}\tilde{K}_\theta(\mathbf{r}_\perp, z)$. On the second step, conventional CT reconstruction, e.g. in the form of the Filtered Back-Projection (FBP) algorithm [5], can be applied to obtain the 3D distribution of the electrostatic potential from its line integrals $\int V_\theta(\mathbf{r}_\perp, z)dz = (\lambda E/\pi)\varphi_\theta(\mathbf{r}_\perp)$. Figure 3(c) shows one example of an axial slice of a lysozyme molecule reconstructed using Eq. (7) from 1800 TEM projections uniformly spaced over 360 degrees, collected at the defocus distance of $z = 45$ Å with monochromatic electron plane wave illumination at $E = 200$ keV. This reconstruction was performed using X-TRACT software [18]. Figure 3(c) demonstrates significantly better reconstruction quality compared to the conventional CT result in Fig. 3(a), which was obtained from the same set of multislice projections. Note also that it is not possible to apply the contrast symmetrization, as done above, to the fully CTF-corrected projections $I(\mathbf{r}_\perp, 0)$, since $K_\pi(\mathbf{q}_\perp^-, 0) = -K(\mathbf{q}_\perp, 0)$, according to Eqs. (4)-(5), and hence $\tilde{K}(\mathbf{q}_\perp, 0) = 0$. This is consistent with Eq. (7) with $z = 0$ and reflects the pure phase nature of the imaged object.

## 4. Conclusions

We have argued that in TEM imaging of small biological molecules or, more generally, in transmission imaging of any "sparsely localized" weakly scattering structures, multiple scattering tends to have only a moderate effect and therefore can be safely ignored in the reconstruction procedures, without introducing large errors into the results. On the other hand, the in-molecule free-space propagation (Fresnel diffraction) cannot be ignored because of the extremely shallow depth of focus under the typical TEM imaging conditions. As DT represents precisely the technique which takes into account the free-space propagation, but not the multiple scattering between different atoms, it appears to be a very good match for this case, as was argued by other authors previously [7,8]. However, acquisition of images on a dense angular grid over the full $2\pi$ range, as required in DT, can often be challenging. As one way to alleviate this problem, in the second part of this work [19] we consider a technique of "pattern matching tomography", that we have called PMT for short, that fully exploits the information about the 3D atom locations as available in TEM defocus series. Such information can be naturally present, for example, in typical cryo-EM data.

**Acknowledgements**

T.E.G. and A.K. acknowledge discussions with Prof. David Paganin related to the present work. T.E.G. is also grateful to Dr. Sheridan Mayo and Mr. Darren Thompson for helpful advices. The authors wish to thank Dr. Andrew Martin for sharing his software code that was used in the course of this research.